\documentclass[conference]{IEEEtran}
\IEEEoverridecommandlockouts
\usepackage{cite}
\usepackage{amsmath,amssymb,amsfonts}
\usepackage{algorithmic}
\usepackage{graphicx}
\usepackage{textcomp}
\usepackage{xcolor}
\usepackage{times}
\usepackage{fancyhdr,graphicx,amsmath,amssymb}
\usepackage[ruled,vlined,linesnumbered]{algorithm2e}
\usepackage{floatrow}  
\include{pythonlisting}
\usepackage{flushend}
\usepackage{float}   
\usepackage{subfigure}
\usepackage{caption}
\usepackage{makecell}

\def\BibTeX{{\rm B\kern-.05em{\sc i\kern-.025em b}\kern-.08em
    T\kern-.1667em\lower.7ex\hbox{E}\kern-.125emX}}  

\begin{document}

\title{Channel Reciprocity Based Attack Detection for Securing UWB Ranging by Autoencoder

}

\author{
\IEEEauthorblockN{\IEEEauthorrefmark{1}Wenlong Gou, \IEEEauthorrefmark{1}Chuanhang Yu, \IEEEauthorrefmark{1}Juntao Ma, \IEEEauthorrefmark{1}Gang Wu, \IEEEauthorrefmark{2}Vladimir Mordachev}
\IEEEauthorblockA{\IEEEauthorrefmark{1} National Key Laboratory of Wireless Communications,\\
University of Electronic Science and Technology of China, Chengdu, China\\
\IEEEauthorrefmark{2} Belarusian State University of Informatics and Radioelectronics}

{\{gouwenlong, chuanhangyu, juntaoma\}}@std.uestc.edu.cn,\\ wugang99@uestc.edu.cn (corresponding author),
mordachev@bsuir.by
}
\maketitle

\begin{abstract}

A variety of ranging threats represented by Ghost Peak attack have raised concerns regarding  the security performance of Ultra-Wide Band (UWB) systems  with the finalization of the IEEE 802.15.4z standard. Based on channel reciprocity, this paper proposes a low complexity attack detection scheme that compares  Channel Impulse Response (CIR) features of both ranging sides utilizing an autoencoder with the capability of data compression and feature extraction.  Taking Ghost Peak attack as an example, this paper demonstrates  the effectiveness, feasibility and generalizability of the proposed attack detection scheme through simulation and experimental validation. The proposed scheme achieves an attack detection success rate of over 99\% and can be implemented  in current  systems at low cost.

\end{abstract}


\section{Introduction}
Ultra-Wideband (UWB) characterized by high resistance to multipath fading{\cite{g1}} and low power consumption offers centimeter-level ranging precision, which has garnered significant
attention in various fields such as keyless car entry and mobile payment{\cite{g24,g4}}. Despite the new release of IEEE 802.15.4z standard in 2020 which has enhanced the accuracy and security of ranging by introducing Scrambled Timestamp Sequence(STS) encrypted by the Advanced Encryption Standard (AES){\cite{g3}} , ranging security remains a critical consideration. 
For the IEEE 802.15.4a protocol, Poturalski et al. {\cite{g9}} proposed the Cicada attack that continuously injects UWB pulses into the receiver during the legitimate transmission of the preamble, in addition to the Early Detection/Late Commitment (ED/LC) attack scheme that leverages the predictability of the signal structure within the preamble{\cite{g10}}. Targeting the IEEE 802.15.4z protocol, the Cicada{++} attack executes the distance attack by transmitting pseudo-random STS signals to alter the timestamps of received signals, whereas the Adaptive Injection Attack (AIA) can further refine the attack precision by controlling the placement of injected attack signals{\cite{g11}}. Patrick Leu et al. {\cite{g12}} conducted the Ghost Peak attack achieving a success rate of up to 4\% on commercially available Apple U1 and Qorvo UWB chips. In \cite{g13}, Claudio Anliker et al. also proposed and demonstrated the Mix-Down attack, which exploits the clock drift of transceivers.
               

To defend against these distance attacks, there are some methods that have been proposed,  e.g., a ranging scheme combining Time of Flight (TOF) and Received Signal Strength (RSS) is proposed in \cite{g14}, aiming to effectively mitigate distance fraud. Additionally, Chen H et al. {\cite{g17}} designed the UnSpoof UWB localization system capable of pinpointing  the position of both the attacker and the legitimate device, and Kiseok Kim et al. {\cite{g19}} proposed a UWB localization system for vehicles based on Directed-Acyclic Graph (DAG) structure to enhance security. However, how to efficiently detect attacks in the process of UWB secure ranging is also an urgent problem that is often neglected.

It is worth noting that attacks such as Ghost Peak require sniffing in advance of the attack, and attacks like Cicada require constant attempts to succeed. 
Therefore, effective detection of attacks is crucial for enhancing the security performance of UWB systems.
In \cite{g15}, Mridula Singh presented a novel modulation technique to detect distance enlargement attacks relying on the interleaving of pulses of different phases. Kyungho Joo et al. {\cite{g16}} achieved an attack detection success rate of 96.24\% by leveraging the consistency of cross-correlation results between the sub-fields of STS and local templates. Towards the upcoming IEEE 802.15.4ab protocol, Li Sun et al. {\cite{u3}} proposed an integrity  protection mechanism based on time-reversal to detect interference, and they also suggested adding a new STS configuration to support integrity protection. Nevertheless, most of the defense schemes against distance attacks modify the established UWB physical layer standards to a large extent, increasing the expenses of practical deployments \cite{c1}.

The UWB channel provides more possibilities for attack detection. Previously, the security and robustness of secret key generation method using UWB channels have been investigated in \cite{g25}. 
Furthermore,  Philipp Peterseil et al. presented a trustworthiness score based on autoencoders trained on Channel Impulse Response (CIR), which could remarkably improve ranging accuracy in {\cite{g26}}.
Inspired by the aforementioned work, this paper proposes an attack detection scheme utilizing channel reciprocity and autoencoders, which could enhance the security performance of the existing UWB system effectively. 
Different from above work, the proposed scheme achieves a low complexity end-to-end attack detection based on existing standards by using only the encoder module of the autoencoder which is trained for CIR feature extraction in the offline training phase. Our main contributions can be summarized as follows:
\begin{itemize}
\item The channel reciprocity in UWB ranging process is analysed, and based on the analysis an attack detection scheme by comparing the CIR of both ranging sides is proposed, which achieves the high reliability of UWB ranging and maintains the current physical layer specification. The feasibility of the proposed scheme is validated through simulation and experiments. 
\item Leveraging the designed autoencoder with a high capacity of data compression trained on simulation data only, the proposed attack detection scheme offers a relatively low cost for transmission, and the generalizability of the scheme is validated through practical deployment. 
\end{itemize}


\section{System Model and problem Formulation}\label{System Model}

\subsection{Classic UWB Ranging Model}

\begin{figure}[htbp]
               
			\centering   
			
			\includegraphics[width=1\linewidth]{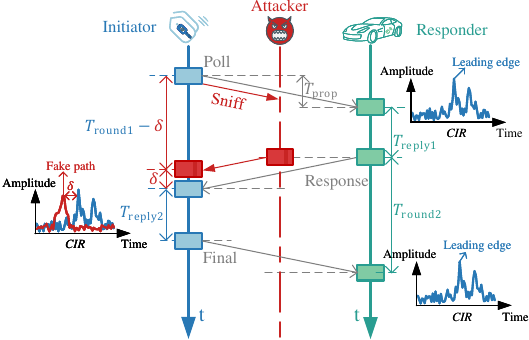}
			\caption{The principle of DS-TWR and Ghost Peak attack}
                \label{fig1}
\end{figure}

Classic UWB ranging methods primarily consist of Single Side-Two Way Ranging (SS-TWR) and Double Side-Two Way Ranging (DS-TWR). Compared with the simple SS-TWR, the DS-TWR method where both the transmitter and receiver exchange a total of three ranging messages can effectively alleviate the impacts of clock drift and other factors{\cite{c2}}. The distance can be estimated as: 



\begin{equation}
\label{eq1}
    d=c\cdot T_{\text{prop}}=\frac{T_{\text{{round}}_\text{1}}\times T_{\text{{round}}_\text{2}}-T_{\text{{ reply}}_\text{1}}\times T_{\text{{ reply}}_\text{2}}}{T_{\text{{ round}}_\text{1}}+T_{\text{{ round}}_\text{2}}+T_{\text{{reply}}_\text{1}}+T_{\text{{ reply}}_\text{2}}}c, 
\end{equation}
where the time intervals $T_{\text{{round}}_\text{1}}$, $T_{\text{{round}}_\text{2}}$, $T_{\text{{ reply}}_\text{1}}$, $T_{\text{{ reply}}_\text{2}}$ are shown in Fig.~\ref{fig1}, $c$ refers to the speed of light.

The reception timestamps used for calculating the time intervals are obtained through STS cross-correlation:
\begin{equation}
    t_i=g_{\text{LE}}(\boldsymbol{s_{\text{STS}}} ,\boldsymbol{s'_{\text{STS}}}), i\in \mathbb{M},   
\end{equation}
where the optional message set $\mathbb{M}$ is $\{\text{Poll}, \text{Response}, \text{Final}\}$,  $\boldsymbol{s_{\text{STS}}}$ and $\boldsymbol{s'_{\text{STS}}}$ denote loacl STS and received STS respectively, $g_{\text{LE}}(\cdot ,\cdot)$ denotes the leading edge detection algorithm based on the Back-Search Time Window (BTW) {\cite{g11}}.  
This detection algorithm is primarily determined by two thresholds: the Maximum Peak to Early Peak Ratio (MPEP) , indicating the ratio between the main path and the first path, and the Peak to Average Power Ratio (PAPR), representing the ratio between peak and average power. Only when both thresholds are satisfied can it be identified as the first path.  
    

The attack detection scheme proposed in this paper is suitable for a wide range of distance attacks, with the Ghost Peak attack used as an illustrative example. 
During the reception of the Response or Final message of the legitimate devices, the attacker transmits an attack signal in which the STS segment, whose power is significantly higher than the legitimate signal, is forged by the attacker. 
These attack signals can alter the timestamps $t_i$ obtained through the leading edge detection algorithm $g_{\text{LE}}$, resulting  in  a shortened measured distance $d${\cite{g12}}. 
The primary process of the Ghost Peak attack is depicted in Fig.~\ref{fig1} (taking the attack on the Response as an example).

\subsection{Improved Integrity check Ranging Model for security }


The absence of a detection step for intentional attacks in existing UWB ranging methods makes both ranging sides continue the ranging process completely all the time even if the measured results are obviously wrong (e.g. excessively drastic changes), which also increases the risk of successful attacks and system power consumption. 
We improve the existing model to enhance its resistance to attacks by adding two modules, i.e., feature extraction and integrity check.

\begin{figure}[htbp]              
    \centering   
    \centerline{\includegraphics[width=0.85\linewidth]{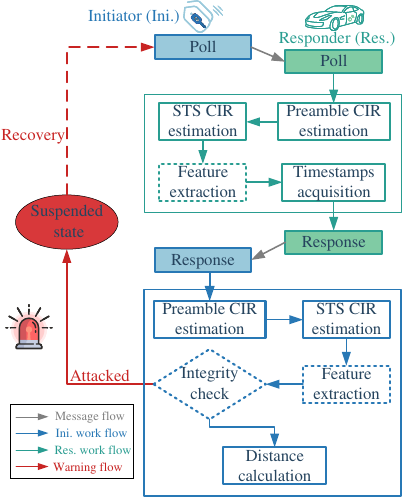}}
    \caption{Improved integrity check ranging model}
    \label{fig3}

\end{figure}

As shown in Fig.~\ref{fig3}, after the Responder receives the first Poll message, it performs CIR estimation using preamble and STS respectively:
\begin{equation}
    \boldsymbol{h_{j}}=g_{\text{CIR}} (\boldsymbol{s_{j}}, \boldsymbol{s'_{j}}), j\in \mathbb{S},
\end{equation}
where the optional sequence set $\mathbb{S}$ is \{\text{preamble}, \text{STS}\}, $g_{\text{CIR}}(\cdot ,\cdot)$ represents CIR estimation function in transceivers, $s_{j}$ and $s'_{j}$ denote local templates and received signals.  
Here we add the feature extraction step utilizing the precise CIR $\boldsymbol{h_{\text{STS}}}$ generated from STS segments. 
In addition, the Initiator also captures the channel features after receiving the second ranging message. Furthermore, we design an integrity checking mechanism to determine the state of the system (i.e. Normal or Attacked) at that time  as a basis to decide whether the ranging process should continue or not. 
The system judged to be under attack will sound an alarm and enter the suspended state, in which the Initiator decides whether to continue ranging.

\subsection{Problem Formulation}
Depending on the application scenario, the implementer of the attack detection can be either the Initiator or Responder during the ranging process. 
Given an observation $\boldsymbol{h^m_{\text{STS}}}\in \Gamma$ $(m\in \{\text{Initiator}, \text{Responder}\})$  over the observation space $\Gamma$ of both ranging sides, attack detection can be regarded as a binary hypothesis testing problem, in which the observations may be derived from two possible hypotheses (Normal or Attacked), denoted by $H_0$ and $H_1$. 
The decision $\psi(\boldsymbol{h^m_{\text{STS}}})$ is made based on the partition of $\Gamma$:
\begin{equation}
    \psi{(\boldsymbol{h^m_{\text{STS}}})} = \left\{ \begin{matrix}
{D_{1},~\boldsymbol{h^m_{\text{STS}}} \in R_{0}~} \\
{D_{2},~\boldsymbol{h^m_{\text{STS}}} \in R_{1}},
\end{matrix} \right.
\end{equation}
where $D_1$ and $D_2$ denote the decision of attacked and normal respectively, while $R_0$ and $R_1$ denote the partition area of $\Gamma$. 
The design of the attack detection aims to find an efficient partition of observation space $\Gamma$. 

The evaluation metrics of attack detection scheme are generally the probability of false alarm $P_{\text{fa}}$ and miss detection $P_\text{m}$, which represent the conditional probabilities $p(D_1|H_0)$ and $p(D_0|H_1)$, respectively.

\section{Attack Detection Scheme Design}
\label{Attack Detection Scheme Design}


\subsection{Attack Detection Procedure Based on Channel Reciprocity}
Channel reciprocity refers to the fact that when both communication ends carry out signal transmission, the transmission characteristics of the channel remain consistent within a specified time window, regardless of the direction of transmission. During a complete DS-TWR ranging process, it is assumed that the Poll message from the Initiator and the Response message from the Responder experience the identical channel fading, which also reflects the assumption of ToF invariance.

\renewcommand{\algorithmcfname}{Procedure}
\begin{algorithm}[!h]
	\SetAlgoLined
	\caption{Attack detection}
	\label{alg:algorithm1}
	\LinesNumbered 
	\KwIn{$f_\varphi\left(\bullet\right)$,$Q_q\left(\bullet\right)$}
	\KwOut{state}
	\While{Receive Poll}{
		Gain $\boldsymbol{h_{\text{IR}}[n]}$\;
		Calculate $\boldsymbol{p}_{\text{IR}}$ via (7)\;
		Embed $\boldsymbol{p}_{\text{IR}}$ in Payload of Response\;
		Send Response\;
	}
	\While{Receive Response}{
		Gain $\boldsymbol{h_{\text{RI}}[n]}$,$\boldsymbol{h_{\text{IR}}[n]}$\;
		Calculate $\boldsymbol{p}_{\text{RI}}$ via (8)\;
		\For {k=1,2,...,$K_{p}$} {
			Compare ${{p}_{\text{IR}}^k}$ and ${{p}_{\text{RI}}^k}$\;
			\If{${p}_{\emph{IR}}^k$$\neq$${p}_{\emph{RI}}^k$}{$d=d+1$}
		}
		\eIf{$d\left({\boldsymbol{p}_{\emph{IR}},\boldsymbol{p}_{\emph{RI}}}\right) \geq T$}
		{state= Attacked}{state= Normal}		
	}
	
\end{algorithm}
In the standard ranging process, both the Initiator and Responder have the process of channel estimation, i.e., both sides involve Channel Impulse Response (CIR).
Utilize the autoencoder to decrease the data dimension of the CIR on both transmitter and receiver sides, mapping the lengthy data vector into a concise feature vector. The mapping process is:
\begin{equation}
\label{eq5}
 \boldsymbol{s}=f_\varphi(\boldsymbol{h[n]}),
\end{equation}
where $\varphi$ denotes the coefficients of the encoder module in the autoencoder and $\boldsymbol{h[n]}$ denotes the input CIR sequence.

Further quantify the extracted low-dimensional feature:
\begin{equation}
\label{eq6}
    \boldsymbol{p}=Q_q\left(\boldsymbol{s}\right),
\end{equation}
where $Q(\cdot)$ denotes the quantization process and $q$ denotes the quantity of quantization bits.

Without loss of generality, it is assumed that the attacker attacks the Response message. In a complete DS-TWR process, the Responder estimates the CIR $\boldsymbol{h_{\text{IR}}[n]}$ at this time after receiving the Poll message. Subsequently, the Responder sends $\boldsymbol{h_{\text{IR}}[n]}$ to the autoencoder for encoding, followed by quantization to obtain $\boldsymbol{p}_{\text{IR}}$:  
\begin{equation}
\label{eq7}
\boldsymbol{p}_{\text{IR}}=Q_q(f_\varphi(\boldsymbol{h_{\text{IR}}\left[n\right]})).
\end{equation}

The Response message containing $\boldsymbol{{p}_{\text{IR}}}$ is then transmitted by the Responder within the Payload field{\cite{g3}}. Upon receiving the Response message, the Initiator demodulates the Payload field to retrieve $\boldsymbol{{p}_{\text{IR}}}$ as well as performs similar operations as the Responder to acquire $\boldsymbol{{p}_{\text{RI}}}$: 
\begin{equation}
\label{eq8}
\boldsymbol{p}_{\text{RI}}=Q_q(f_\varphi(\boldsymbol{h_{\text{RI}}\left[n\right]})). 
\end{equation}


Then each bit ${p}_{\text{IR}}^k$ and ${p}_{\text{RI}}^k$ are individually compared to calculate the Hamming distance between $\boldsymbol{{p}_{\text{IR}}}$ and $\boldsymbol{{p}_{\text{RI}}}$. Ultimately, by comparing this distance with the preset judgment threshold ${T}$ (which can be established through simulation or practical measurement), the attacks can be detected:
\begin{equation}
\label{eq9}
\left\{ {\begin{matrix}
{d( {\boldsymbol{p}_{\text{IR}},\boldsymbol{p}_{\text{RI}}})\geq {T},\text{Attacked state}} \\
{d( {\boldsymbol{p}_{\text{IR}},\boldsymbol{p}_{\text{RI}}})\textless{T},\text{Normal state}},
\end{matrix}}\right. 
\end{equation}
where $d(\cdot,\cdot)$ denotes the Hamming distance. 
The observation space $\Gamma$ is partitioned into two regions through (\ref{eq9}).
The process of the proposed attack detection scheme is summarized in Procedure \ref{alg:algorithm1}. 

\subsection{Attack Detection Principles Using Autoencoders}

UWB Channel features such as Received Signal Power to First Path Power level ratio, mean excess delay spread, kurtosis, etc., can be directly used to identify the Line of Sight (LOS) and Non-Line of Sight (NLOS) channel{\cite{u1}}. However, in order to prevent attackers from using these obvious features to pass the integrity check and to extract the most representative features of the channel, this paper selects neural networks for feature extraction. 

\begin{figure}[tbp]              
    \centering   
    \centerline{\includegraphics[width=1\linewidth]{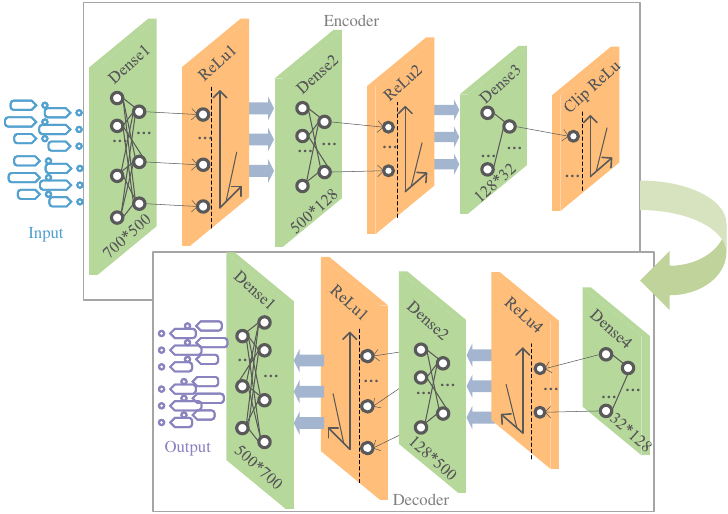}}
    \caption{The structure of designed autoncoder  }
    \label{fig4}
    
\end{figure}

The autoencoder is one of the most widely used self-supervised neural network structures for data compression and feature extraction,
which can extract the most critical features by reproducing the original data{\cite{g22}}.  
To reduce transmission consumption, We use the autoencoder to encode the CIR to extract its low-dimensional features and the CIRs of both sides exhibit similar characteristics due to the consistency of the channel. Meanwhile, since the attacker's signal reaches the receiver through a distinct channel, the attack can be detected by contrasting the CIR features of both legitimate sides. Thus, the autoencoder is only fed by the CIRs in the unattacked situation for training.

The structure of the Multilayer Perceptron (MLP) autoencoder designed in this paper is shown in Fig.~\ref{fig4}.
On the other hand, it is essential to reasonably select the number of encoded feature dimensions and the corresponding quantization bits in order to facilitate transmission. 

\subsection{Complexity Analysis}

The complexity of the proposed attack detection scheme is primarily determined by the autoencoder. Therefore, this subsection concentrates on analyzing the complexity of the designed autoencoder, including time complexity and space complexity. Time complexity is quantified by the count of Floating-Point Operations (FLOPs) and space complexity is measured by the aggregate count of model parameters. 

The time complexity $c_{\text{time}}$ and space complexity $c_{\text{space}}$ of the autoencoder can be expressed as:
\begin{subequations}
\begin{align}
c_{\text{time}}&=\sum_{d=1}^{D}{\left(2I_d-1\right)\times\ J_d}+\sum_{d=1}^{D-1}J_d, \label{Za}\\
c_{\text{space}}&=\sum_{d=1}^{D}{\left(I_d+1\right)\times\ J_d}, \label{Zb}
\end{align}
\end{subequations}
where $D$ denotes the depth of the autoencoder, i.e., the quantity of fully connected layers, $I_d$ and $J_d$ denote the input and output dimensions of each fully connected layer, respectively.


\section{Numerical Simulation Evaluation and Experimental Validation}\label{Numerical Simulation Evaluation and Experimental Validation}

This section shows the simulation evaluation results of the proposed attack detection scheme, and we also conducted practical validation of its effectiveness using the commercial chip DW3110. 
The parameter configurations for the legitimate signal and attack signals are detailed in Table \ref{tab1}.
\begin{table}[htbp]
        \caption{Configurations for legitimate and attack signals}
        \label{tab1}
        \small
	\centering
	\begin{tabular}{c|c|c}
        \hline               
			Parameter&legitimate signals&attack signals\\      
	\hline              
			mode &BPRF&BPRF \\
			Preamble spreading factor  &4&9\\
                Preamble code index &	9&	9\\
                SFD number &	0&	0\\
                modulation	&BPSK+BPM&	BPSK+BPM\\
                Payload encoding & \makecell[c]{RS\\\&convolution}	&\makecell[c]{RS\\\&convolution}\\
                samples of per pulse&	4	&4\\
                Preamble duration&	64	&64\\
                STS segment length	&64&	64\\
	\hline       
        \end{tabular}
\end{table}
\begin{figure}[htbp]              
    \centering 
    \subfigure[Different input dimensions]{
    \includegraphics[width=0.5\linewidth]{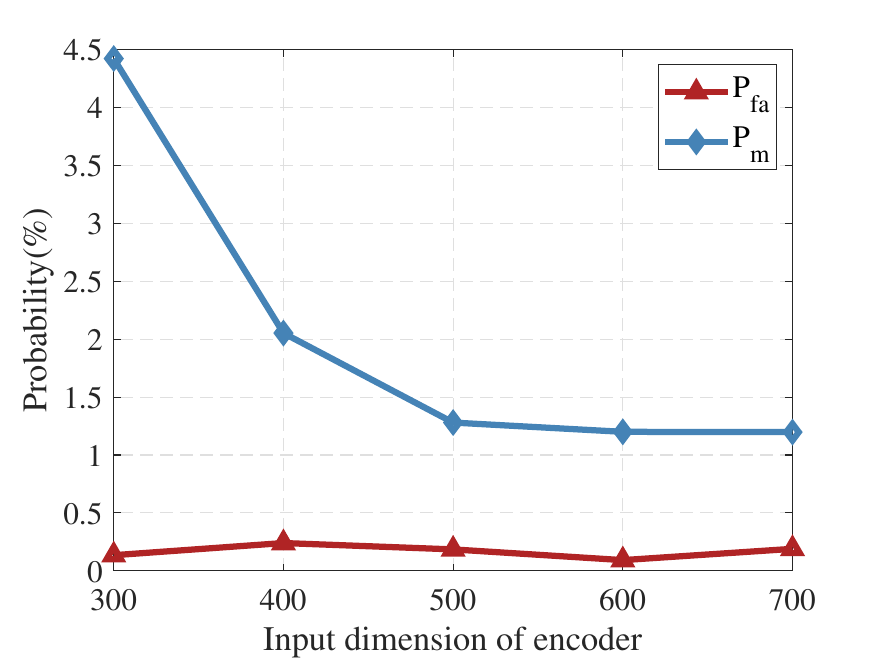}
    \label{fig6a}
    }
    \hspace{-7.5mm}
    \subfigure[Different output dimensions]{
    \includegraphics[width=0.5\linewidth]{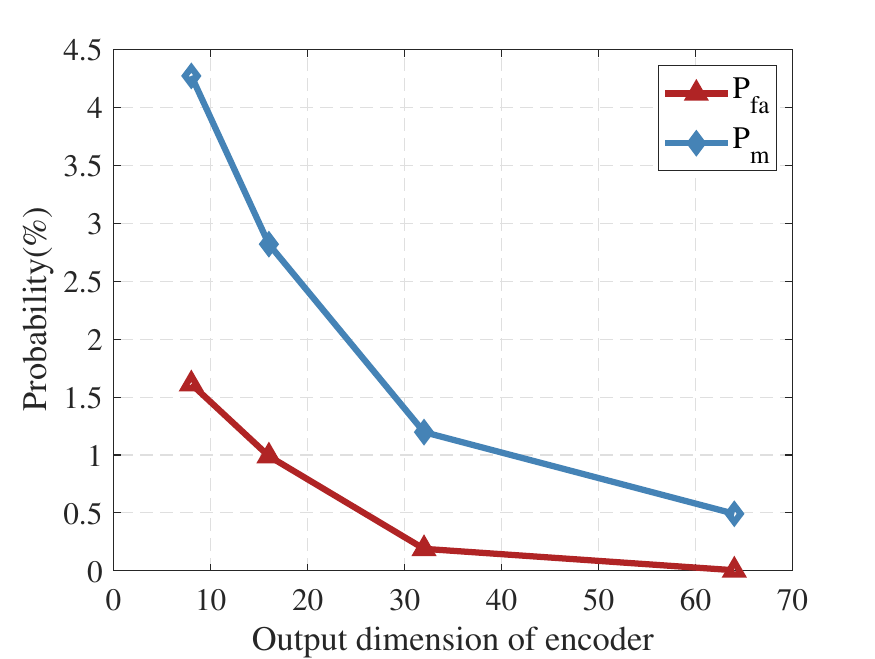}
    \label{fig6b}
    }
    \quad
    \subfigure[Different quantization bits]{
    \includegraphics[width=0.5\linewidth]{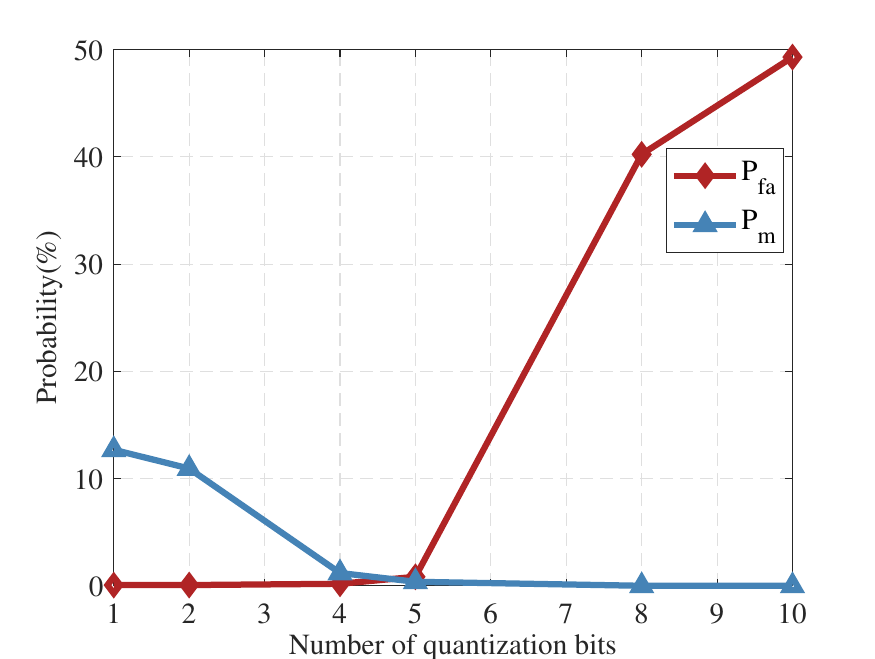}
    \label{fig6c}
    }
    \hspace{-7.5mm}
    \subfigure[Different threshold coefficients]{
    \includegraphics[width=0.5\linewidth]{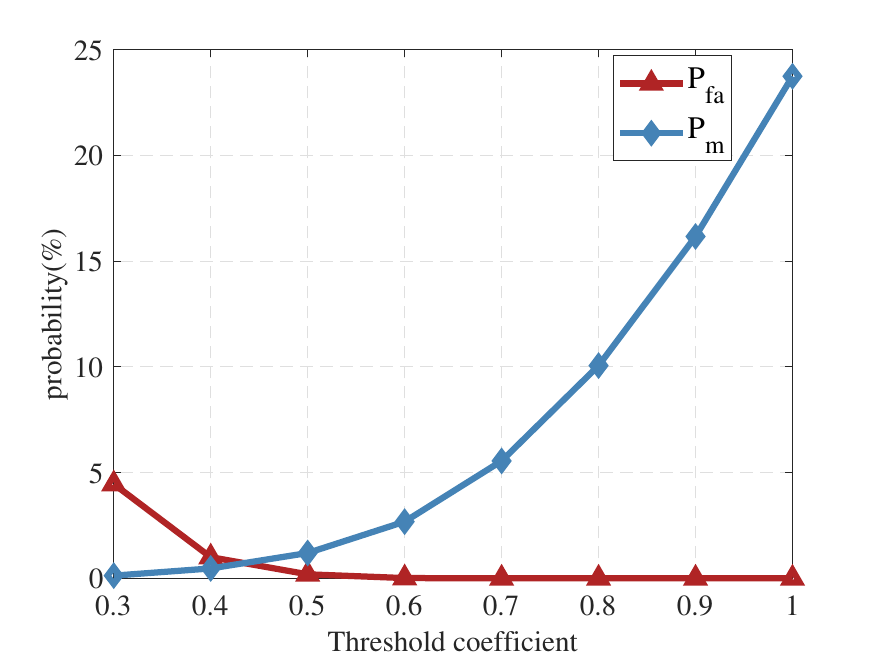}
    \label{fig6d}
    }
    \caption{Performance comparison for different parameters of autoencoder} 
    \label{fig6}
\end{figure}

\begin{figure*}[htbp]
	\centering
        \subfigure[Success probability of attack]{
		\centering
		\includegraphics[width=0.34\linewidth]{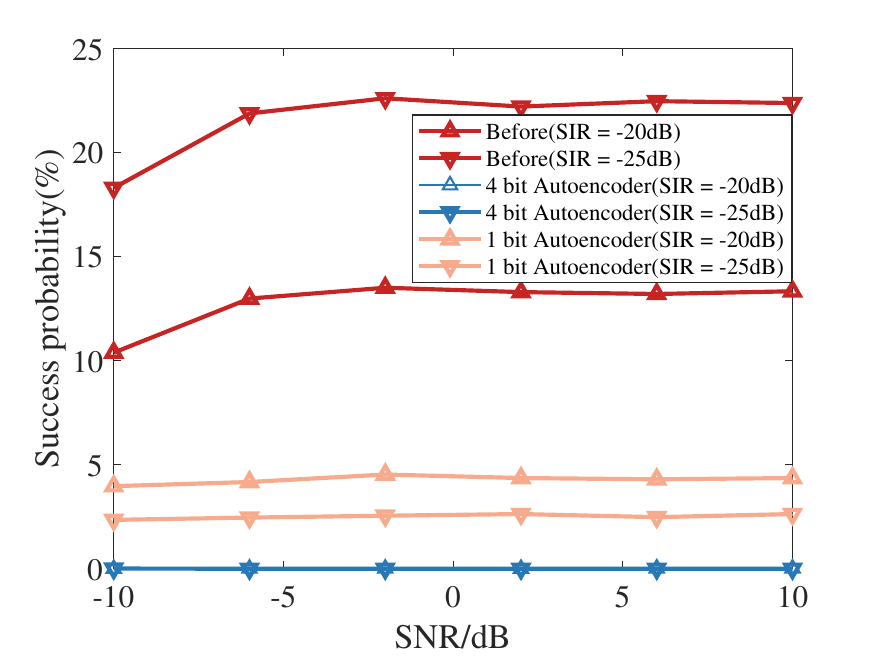}
		\label{fig7a}
  }
  \hspace{-8mm}
	 \subfigure[Probability distribution curve of ranging error]{
		\centering
		\includegraphics[width=0.34\linewidth]{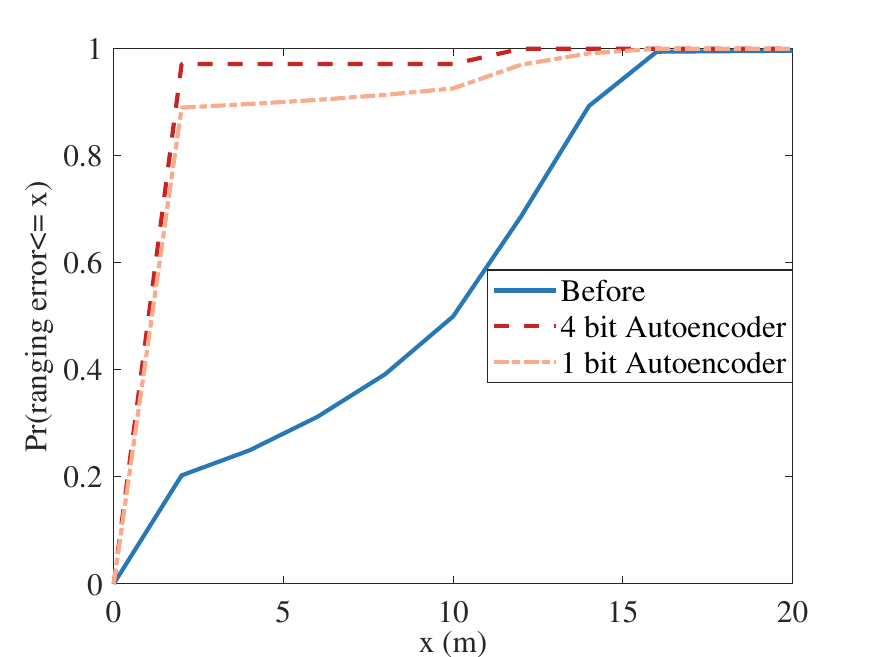}
		\label{fig7b}
  }
  \hspace{-8mm}
   \subfigure[False alarm probability $P_{fa}$]{
		\centering
		\includegraphics[width=0.34\linewidth]{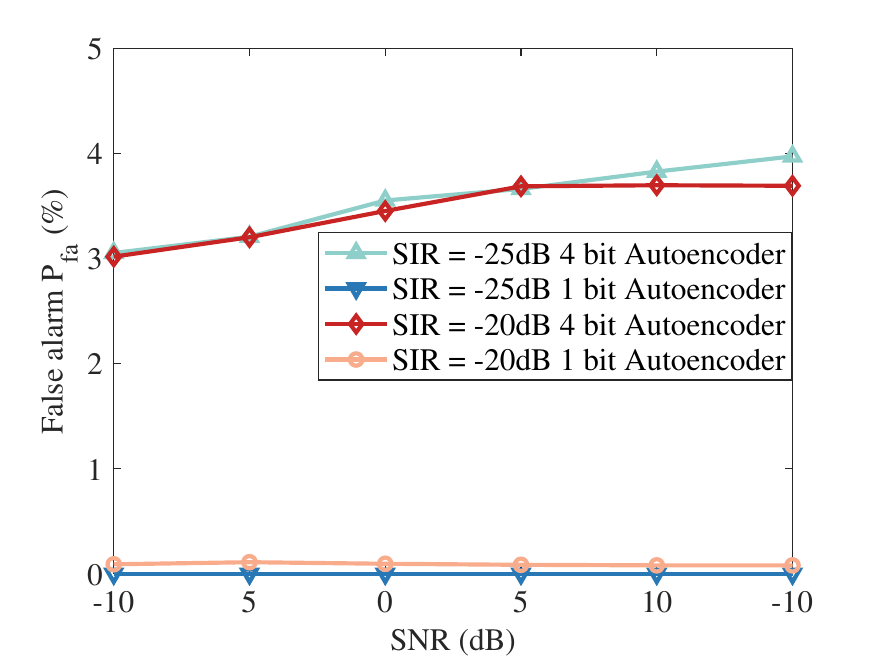}
		\label{fig7c}
  }
  \caption{Performance of attack detection scheme proposed}
  \label{fig7}
\end{figure*}


The receiver structure is not explicitly specified in the IEEE 802.15.4z standard. In this paper, after shaping filtering and sampling on the received signal, the local SHR and STS fields are employed to correlate with the processed signal to search for the first path using the leading-edge detection algorithm described in Section \ref{System Model}. The length of BTW is fixed at 400 samples (with a sampling rate of 2 GHz), MPEP is set to 0.5 and PAPR is set to 2. The RAKE receiver architecture is employed for receiving the PHR and Payload fields. What's more, all simulations in this paper are performed with a distance of 10 m between the devices and an attack is considered successful if the ranging result is below 5 m.

\subsection{Numerical Simulation Evaluation}

Let $\alpha_\text{t}$ denote the threshold coefficient ($0<\alpha_\text{t}<1$) and $T_\text{H}$ denote the Hamming distance between the quantified CIR features of both sides without attacks, then they are related to the preset judgment threshold $T$ as follows: 
\begin{equation}
    T=\alpha_\text{t}\cdot T_\text{H}
\end{equation}



Adopting the indoor LOS channel model specified in the IEEE 802.15.4 standard, the CIRs of both sides in the normal ranging process (a total of 20,000 pairs) are divided into a training set and a test set at a ratio of 9:1. Through simulation analysis, the input dimension, output dimension, number of quantization bits and threshold coefficient in the attack detection scheme finalized in this paper are 700, 32, 4 and 0.5, respectively. 


Fig.~\ref{fig6} illustrates the impact of various parameters on the performance of attack detection. It shows that the input dimension has a minor effect on the false alarm probability, whereas the miss detection probability decreases significantly as the input dimension increases. There is a positive correlation between the output dimension and performance, aligning with the anticipated outcome that a higher number of output features facilitates differentiation between CIRs of normal and attacked state. Nevertheless, larger output dimensions imply higher costs and an elevated risk of overfitting, which requires a trade-off between performance and complexity. 

As shown in Fig.~\ref{fig6c}, a gradual rise in false alarm probability and a decline in miss detection probability are observed with the increase of quantization bits. A higher number of quantization bits lead to more total bits of extracted feature data, which results in the Hamming distance between $\boldsymbol{p}_{\text{IR}}$ and $\boldsymbol{p}_{\text{RI}}$ is more likely to be larger than the preset threshold. 
Fig.~\ref{fig6d} illustrates how the threshold impacts detection performance, aligning with the general knowledge that raising the threshold decreases false alarm probability but increases miss detection probability.

Furthermore, the time and space complexity of the designed scheme are  1,670,692 FLOPs and 838,180 parameters, as calculated from (\ref{Za}) 
(\ref{Zb}), respectively.

Adopting the proposed attack detection scheme, we simulated 10,000 DS-TWR ranging processes under Ghost Peak attack with various parameter configurations. We conducted statistical calculations for the success probability of attack $P_\text{s}$, the probability distribution of ranging error $P_\text{r}$ and false alarm probability $P_{\text{fa}}$. The obtained results are shown in Fig.~\ref{fig7}. The SIR in Fig.~\ref{fig7} represents the ratio of STS power of the legitimate signal to the attack signal, while SNR denotes the signal-to-noise ratio in the environment. The results indicate that the proposed scheme can effectively detect attacks. Fig.~\ref{fig7a} shows that the success probability of attack detection can exceed 99\% with 4 quantization bits and surpass 95\% with 2 quantization bits. Fig.~\ref{fig7b} illustrates the probability distribution curve of ranging error, showing that the proposed scheme can effectively reduce ranging error in a large extent, and the result when using 4-bit quantization is slightly better than 1-bit quantization. Fig.~\ref{fig7c} depicts the curve of false alarm probability with respect to SNR. Whereas the false alarm probability with SNR has little difference, it tends to increase to some extent as the power of the attack signal decreases. This may be attributed to the insufficient power of the attack signal, resulting in an insufficient distinction in the CIR between both sides, thereby increasing the false alarm probability.

 

\subsection{Experimental Validation}

\begin{figure}[htbp]              
    \centering   
    \centerline{\includegraphics[width=1\linewidth,height=1\linewidth]{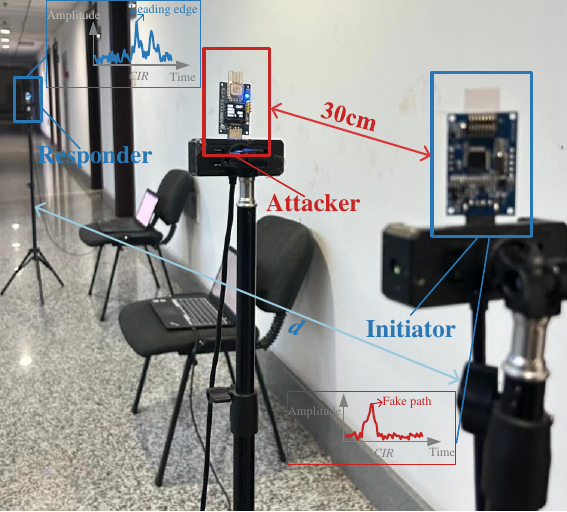}}
    \caption{Practical test environment}
    \label{fig8}
\end{figure}
We utilized the commercial ranging device equipped with the DW3110 chip to build a Ghost Peak attack scenario. The attack detection model,  obtained by offline training using simulation data, was practically deployed to validate the proposed scheme in an indoor 10m test environment, as shown in Fig.~\ref{fig8}. The STS mode of the chip was switched to the SDC mode which is more likely to be attacked successfully{\cite{g23}}, with the statistically measured results  shown in Table \ref{tab2}.

\begin{table}[htbp]
        \caption{The result of practical validation}
        \label{tab2}
        \small
	\centering
	\begin{tabular}{c|c|c|c}
        \hline  
         \textbf{Befor detection}&\multicolumn{3}{c}{\textbf{After detection}}\\
         \hline
			 \makecell[c]{$P_\text{s}$}   &$P_{\text{fa}}$&$P_\text{m}$&$P_\text{s}$ \\ 
    \hline 
   			60.067\% &2.4\%&0.075\% &0.045\%\\
	\hline              

        \end{tabular}
\end{table}

In terms of listed results, it is evident that the proposed attack detection scheme performs satisfactorily in the practical scenario. The proposed scheme is able to detect attacks with a successful probability of nearly 99\%  while maintaining a low false alarm probability, which further validates the feasibility and generalizability of the designed model.

\section{Conclusion}\label{Conclusion}

Leveraging the principle of channel reciprocity and an autoencoder with capability of data compression and feature extraction, this paper proposes an attack detection scheme that compares the CIR characteristics of both sides. The transmission consumption is greatly reduced by incorporating the quantization process. In the meanwhile, the scheme is able to be relatively compatible with existing UWB systems by means of offline training and online deployment. We also evaluated and validated the effectiveness of the scheme through simulation and practical experiments. 
Furthermore, other factors (e.g. channel variation and  threshold update) can be jointly considered to optimise the performance of the proposed scheme in future work.


\vspace{12pt}

\end{document}